\pdfoutput=1

\documentclass[%
 reprint,
 floatfix,
 widetext,
 amsmath,amssymb,
 aps,
]{revtex4-2}

\usepackage[strict]{changepage}
\usepackage[title]{appendix}
\usepackage{xcolor}
\usepackage{graphicx}
\usepackage{dcolumn}
\usepackage{bm}

\usepackage[utf8]{inputenc}
\usepackage{physics}
\usepackage{amsfonts}
\usepackage{amsmath}
\usepackage{xcolor}
\usepackage{array}
\newcolumntype{M}[1]{>{\centering\arraybackslash}m{#1}}

\begin{document}

\preprint{APS/123-QED}

\title{Robust correlated magnetic moments in end-modified graphene nanoribbons}

\author{Antoine Honet
}
\altaffiliation{Present address: Department of Electrical Engineering, Eindhoven University of Technology, Eindhoven 5612 AP, Netherlands}
\affiliation{%
Department of Physics, Namur Institute of Structured Materials, University of Namur, Rue de Bruxelles 51, 5000 Namur, Belgium
}%

\author{Luc Henrard}
\affiliation{%
Department of Physics, Namur Institute of Structured Materials, University of Namur, Rue de Bruxelles 51, 5000 Namur, Belgium
}%

\author{Vincent Meunier}%
\affiliation{%
Department of Engineering Science and Mechanics, The Pennsylvania State University, State College, PA, USA
}%

\date{\today}

\begin{abstract}
We conduct a theoretical examination of the electronic and magnetic characteristics of end-modified 7-atom wide armchair graphene nanoribbons (AGNRs). Our investigation is performed within the framework of a single-band Hubbard model, beyond a mean-field approximation. First, we carry out a comprehensive comparison of various approaches for accommodating di-hydrogenation configurations at the AGNR ends. We demonstrate that the application of an on-site potential to the modified carbon atom, coupled with the addition of an electron, replicates phenomena such as the experimentally observed reduction in the bulk-states (BS) gap. These results for the density of states (DOS) and electronic densities align closely with those obtained through a method explicitly designed to account for the orbital properties of hydrogen atoms. Furthermore, our study enables a clear differentiation between mean-field (MF) magnetic moments, which are spatially confined to the same sites as the topological end-states (ES), and correlation-induced magnetic moments, which exhibit localization along all edges of the AGNRs. Notably, we find the robustness of these correlation-induced magnetic moments relative to end modifications, within the scope of the method we employ.
 
\begin{description}
\item[Keywords]
Graphene nanorribons, magnetic moments, correlation, Hubbard model, mean-field approximation, GW approximation, topological end states
\end{description}
\end{abstract}

\maketitle


\author{Antoine Honet, Luc Henrard and Vincent Meunier}

\section{Introduction}

Graphene nanoribbons (GNRs) have been the subject of many studies in the last two decades both theoretically~\cite{son_energy_2006, brey_electronic_2006, yang_quasiparticle_2007, yazyev_emergence_2010, lu_competition_2016, hagymasi_entanglement_2016, honet_2023_effect} and experimentally~\cite{ruffieux_electronic_2012, talirz_termini_2013, zhang_-surface_2015, kimouche_ultra-narrow_2015, sode_electronic_2015, wang_giant_2016, talirz_-surface_2017, talirz_band_2019, lawrence_probing_2020}. This interest in GNRs is explained in part by the possibility of inducing a band gap in graphene nanosystems while extended graphene has a zero band gap~\cite{castro_neto_electronic_2009, son_energy_2006, brey_electronic_2006, yang_quasiparticle_2007}. GNRs are also interesting because, for example, finite-sized armchair graphene nanoribbons (AGNRS) and AGNRS heterojunctions are known to host topological states~\cite{cao_topological_2017, lopez-sancho_topologically_2021, jiang_topology_2021}. AGNRs of different widths can now be synthesized using a bottom-up approach with atomic precision~\cite{ruffieux_electronic_2012, talirz_termini_2013, zhang_-surface_2015, kimouche_ultra-narrow_2015, sode_electronic_2015, wang_giant_2016, talirz_-surface_2017, talirz_band_2019, lawrence_probing_2020}. This allows not only the study of the fundamental properties of specific AGNRs but also the engineering of GNRs with well-defined electronic properties. 

In the process of synthesizing 7-atom-wide AGNRs (7-AGNRs), different possible end terminations have been observed~\cite{talirz_termini_2013}. The influence of termination on the bandgap value was studied in Ref.~\onlinecite{talirz_-surface_2017} both experimentally and theoretically using the density functional theory (DFT) and tight-binding (TB) methods. In that study, the end modifications include dehydrogenation and di-hydrogenation of the central carbon (C) atom at the zigzag ends. It was observed that di-hydrogenation of the two ends leads to the reduction of the bulk-state (BS) bandgap, defined as the bandgap between states that are not topological end-states (ES). This BS bandgap reduction was reproduced using a single-band TB model and removing the C atom sites where the di-hydrogenation took place since they cannot contribute with an electron to the $\pi$-system~\cite{soriano_magnetoresistance_2011, talirz_-surface_2017}. 

Furthermore, doped GNRs can be produced by introducing substituent to C atoms such as nitrogen (N) or boron (B)~\cite{kawai_atomically_2015, cloke_site-specific_2015, kawai_multiple_2018}. It is possible to describe such substitution in the TB framework, adapting the number of electrons and setting an on-site potential at the substituent atomic sites. One electron is added (resp., removed), and the on-site potential is set to a negative (resp., positive) value for a N (resp,. B) substitution~\cite{latil_mesoscopic_2004, khalfoun_b_2010, lambin_long-range_2012, khalfoun_long-range_2014, joucken_charge_2015}.

Magnetic moments in graphene nanostructures are important for technological applications. They are often studied using a mean-field (MF) approximation of the Hubbard model~\cite{fernandez-rossier_magnetism_2007, kumazaki_tight-binding_2008, yazyev_emergence_2010, soriano_magnetoresistance_2011, pike_tight-binding_2014, mishra_topological_2020}. When electron-electron effects are included, a correlation part has to be included in the magnetic moment expression, which accounts for the non-decoupling of double occupancies~\cite{feldner_magnetism_2010, raczkowski_interplay_2017, joost_correlated_2019, raczkowski_hubbard_2020, honet_exact_2022}. The relation between topological states energy renormalization and local magnetic moments was recently investigated in GNR heterojunctions using a many-body GW approximation for inclusion of correlation effects~\cite{joost_correlated_2019}. In this reference, it was shown that the magnetic moments in MF are predicted to be spatially localized exactly where the zero-energy states are located while they are larger in the GW approximation with a larger range of values in the system. Moreover, they are located along all edges of the GNRs and not only at the location of the zero-energy states. Because magnetic moments are strongly affected by correlation, we study them in this article in pristine and end-modified 7-AGNRs. We investigate the spatial localization of the magnetic moments in MF and GW approximation by changing the number of electrons and the on-site potential at the modified atomic sites. 

The rest of this paper is organized as follows: we start by reviewing the models and methods used throughout this study in section~\ref{sec:models_methods}. We then compare in more details different ways of modelling di-hydrogenation within the Hubbard model framework in section~\ref{sec:different_modellings}. Next, we adopt a common model for all end-modification scenarios to study the magnetic moments induced by topological ES and by correlation effects in section~\ref{sec:topol_correl_magn_mom}. We also study the robustness of these magnetic moments against end-modifications of the AGNRs, contrasting them with topologically-induced and correlated magnetic moments.

\section{Models and methods}
\label{sec:models_methods}

\subsection{Single-band Hubbard model for extended graphene and nanoflakes with edges passivated with hydrogen atoms}

In extended graphene, each C atom is bound to three other C atoms, leading to $sp^2$ hybridization. As a result, each C atom contributes one $p_z$ electron to the $\pi$ system, allowing the use of the single-band TB or Hubbard model. In the case of graphene nanoflakes, such as finite-size AGNRs, single-band models can be used if one assumes that each C atom at the edges is passivated by exactly one H atom. The C atoms at the edges are then bound to two other C atoms and one H atom, leading to $sp^2$ hybridization. We therefore model these systems with TB or Hubbard Hamiltonians at half-filling, \textit{i.e.}, with the number of electrons being equal to the number of C atoms.

The single-band TB Hamiltonian containing only nearest-neighbor hopping terms reads:
\begin{equation}
\hat{H}_{TB } = \sum_{i,\sigma} \epsilon_{i,\sigma} \hat{c}^\dagger_{i\sigma} \hat{c}_{i\sigma}  - t \sum_{<ij>, \sigma}  (\hat{c}^\dagger_{i\sigma} \hat{c}_{j\sigma}  +c.c.), 
\label{eq:TB_ham}
\end{equation}
where $\epsilon_{i,\sigma}$ are the on-site potentials, $t$ is the hopping parameter, $\hat{c}^\dagger_{i\sigma}$ (resp. $\hat{c}_{i\sigma}$) is a creation (resp. destruction) operator of an electron at atomic site $i$ with spin $\sigma$. The $\langle$ $\rangle$ sign under the summation symbol indicates that the sum runs over all pairs of nearest neighbors. In the event that all atoms are equivalent, as assumed in pure carbon systems, all on-site potentials are equal and they only lead to a global shift in energy. We therefore arbitrarily set them to zero. Typical values for the hopping parameter in graphene are around or slightly below $3 \hspace{0.1 cm}$eV~\cite{yazyev_emergence_2010, castro_neto_electronic_2009, honet_semi-empirical_2021} and we took $t=2.7 \hspace{0.1cm}$ eV throughout this work.
 
The single-band Hubbard model is obtained by adding an interaction term proportional to the interaction parameter $U$:
\begin{equation}
\hat{H}_{Hubbard } = \hat{H}_{TB } + U \sum_i \hat{n}_{i \uparrow} \hat{n}_{i \downarrow},
\label{eq:Hubbard_ham}
\end{equation}
where $\hat{n}_{i \sigma} = \hat{c}^\dagger_{i\sigma} \hat{c}_{i\sigma}$ is the density operator (of electrons on atomic site $i$ and with spin $\sigma$). In this paper we used $U=2t$, which is a typical realistic value for carbon nanostructures~\cite{joost_correlated_2019, honet_semi-empirical_2021, honet_2023_effect}.

\subsection{Modelling N or B substitutions}

Starting the TB Hamiltonian of pure C systems described in eq.~(\ref{eq:TB_ham}), one can model the substitution of one C atom by an N or a B atom by changing the on-site potential $\epsilon$ at the substitutional site and by changing the number of electrons. Since N atoms have one electron more than C atoms, one electron is added in the $\pi$ system for each N substitution. The on-site potential for the N atoms is set to a negative value of several eV~\cite{latil_mesoscopic_2004, khalfoun_b_2010, lambin_long-range_2012, khalfoun_long-range_2014, joucken_charge_2015}, meaning that it attracts more electrons than the other C atoms, accounting for the different atomic numbers. The case of substitution for B is modeled analogously by removing one electron from the $\pi$ system and setting a positive on-site value of several eV~\cite{latil_mesoscopic_2004, khalfoun_b_2010, lambin_long-range_2012, khalfoun_long-range_2014, joucken_charge_2015}. The resulting model Hamiltonian reads:
\begin{equation}
\begin{split}
\hat{H}_{subst } = \epsilon_{N/B} \sum_{\alpha \in subst,\sigma}  & \hat{c}^\dagger_{\alpha \sigma} \hat{c}_{\alpha \sigma}  - t \sum_{<ij>, \sigma}  (\hat{c}^\dagger_{i\sigma} \hat{c}_{j\sigma}  +c.c.) \\
& + U \sum_i \hat{n}_{i \uparrow} \hat{n}_{i \downarrow},
\end{split}
\label{eq:subst_ham}
\end{equation}
where the $\alpha$ index runs over all substitutional sites and the $\epsilon_{N/B}$ is the N or B on-site potential.

\subsection{Modelling di-hydrogenation}
\label{sec:methods_model_di_hydro}

There exist several ways to model the effect of di-hydrogenation on a given C atomic site in graphene. The first one, called \textit{C-removing}, consists in removing the affected sites of the TB / Hubbard model~\cite{talirz_-surface_2017, soriano_magnetoresistance_2011}. The electron that the C atom shares with the $\pi$ system in the single-hydrogen passivated case is now used to bind to the second hydrogen. The Hamiltonian operators in this case are simply the ones described by eqs.~(\ref{eq:TB_ham}) or~(\ref{eq:Hubbard_ham}) where the sums run over the C atomic sites except the ones where di-hydrogenation occurs. The number of electrons remains equal to the number of C atoms, \textit{i.e.} the models are considered at half-filling.

A second way of modelling di-hydrogenation is by considering H atomic sites as potential atomic sites for the electron to be localized and therefore by adding H orbitals and associated hopping/on-site parameters in the Hamiltonian. Therefore, we refer to this approach using the \textit{H-orbitals} denomination. One H atom comes with one electron such that the total number of electrons is the number of C atomic site plus one for each site of di-hydrogenation. The Hamiltonian accounting for di-hydrogenated in nanographene samples is therefore given by the following equation:
\begin{equation}
\hat{H} =  \hat{H}_{Hubbard}  + \hat{H}_H,
\end{equation}
where $\hat{H}_{Hubbard}$ is given at eq.~(\ref{eq:Hubbard_ham}) and $\hat{H}_H$ is a H-related Hamiltonian given by:
\begin{equation}
\hat{H}_H = \epsilon_H  \sum_{\alpha, \sigma} \hat{h}_{\alpha \sigma}^\dagger \hat{h}_{\alpha \sigma} + t_h \sum_{\alpha, \sigma} ( \hat{h}_{\alpha \sigma}^\dagger \hat{c}_{h_\alpha,\sigma} + c.c.),
\end{equation}
where index $\alpha$ runs over all H atoms added for di-hydrogenation, $\hat{h}_{\alpha \sigma}$ (resp., $\hat{h}_{\alpha \sigma}^\dagger$) is the annihilation (resp., creation) operator of an electron on the H atom labeled $\alpha$ with spin $\sigma$, the notation $\hat{c}_{h_\alpha}$ denotes the annihilation operator on the C atom to which the H label $\alpha$ is adsorbed, $\epsilon_h$ is the on-site parameter at the H site, and $t_h$ is the hopping parameter linking the H atom and the C atom where H is added. These parameters were chosen to be $\epsilon_h=-t/16$ and $t_h=2t$~\cite{wehling_resonant_2010}.

We considered a third way of modelling di-hydrogenation. As previously, the basic idea is that an electron should be added to the system and forced to stay close to the C atom that hosts the di-hydrogenation site. According to these principles, we propose to model di-hydrogenation by adding an electron in the system and keeping the initial system composed of only C atoms, \textit{i.e.} not removing any C sites nor introducing any H sites. Instead, the localization around the C atoms subject to di-hydrogenation is modelled by setting a large negative value for its on-site potential~\cite{kumazaki_tight-binding_2008}. We name this third method \textit{C-on-sites}. The model Hamiltonian thus reads:
\begin{equation}
\begin{split}
\hat{H}_{o-s } = \epsilon_{o-s} \sum_{\alpha \in \{C_H\},\sigma}  & \hat{c}^\dagger_{\alpha \sigma} \hat{c}_{\alpha \sigma}  - t \sum_{<ij>, \sigma}  (\hat{c}^\dagger_{i\sigma} \hat{c}_{j\sigma}  +c.c.) \\
& + U \sum_i \hat{n}_{i \uparrow} \hat{n}_{i \downarrow},
\end{split}
\label{eq:on_site_full_C}
\end{equation}
where the $\alpha$ index runs over all C atomic sites that are subject to di-hydrogenation (the ensemble of these C sites is written $\{C_H\}$.

One can easily see that the Hamiltonians described by  eqs.~(\ref{eq:subst_ham}) and~(\ref{eq:on_site_full_C}) are identical up to the order of magnitude of the on-site potential values. This similarity could allow one to simulate di-hydrogenation and substitution in a unified framework. In section~\ref{sec:different_modellings} we compare the different approaches for modelling di-hydrogenation and we show that the \textit{C-on-sites} method can capture features that are also observed in the two other methods for large enough negative on-site values. This allows us to consider only the \textit{C-on-sites} method for the di-hydrogenation modelling in an attempt to unify the description of end-modifications.

\subsection{MF approximation of the Hubbard term}

The interaction term (second term of eq.~(\ref{eq:Hubbard_ham})) is often treated in a MF approximation to model graphene's electronic properties~\cite{yazyev_emergence_2010, bullard_improved_2015, mishra_collective_2020, mishra_topological_2020}. The MF approximation consists of decoupling the product of two density operators in the interaction term $\hat{n}_{i \uparrow} \hat{n}_{i \downarrow}$. The approximation is then between the density operator of one spin and mean density of the opposite spin:
\begin{equation}
\begin{split}
    \hat{H}_{Hub, MF}  = \hat{H}_{TB}  +\sum_{i} U (\hat{n}_{i\uparrow} \langle \hat{n}_{i\downarrow} \rangle + \langle \hat{n}_{i\uparrow} \rangle \hat{n}_{i\downarrow} ),
\end{split}
\label{eq:Hubbard_MF}
\end{equation}
where $\langle \hat{n}_{i\sigma} \rangle$ is the mean value of the operator $\hat{n}_{i\sigma }$. By adopting such an approximation, the products of deviations with the mean densities and a constant shift in the Hamiltonian are neglected~\cite{honet_mean_field_2023}. The Hamiltonian of eq.~(\ref{eq:Hubbard_MF}) has to be solved self-consistently, starting from an initial guess for the mean densities and updating them at each step, where a new Hamiltonian is diagonalized.

\subsection{GW approximation}

The GW approximation is a beyond-MF approximation that includes some correlation effects \textit{via} dynamically-screened interaction. The approximation was recently applied to the Hubbard model in the context of graphene nanostructures~\cite{joost_correlated_2019, honet_semi-empirical_2021, honet_2023_effect}. The GW approximation is based on Hedin's equations~\cite{hedin_effects_1970} that are approximated according to the vertex function that leads to Dyson's equation:
\begin{equation}
    G^{R} (\omega) = G^{R}_0 (\omega) + G^{R}_0 (\omega) \Sigma^{R} (\omega) G^{R} (\omega),
\label{eq:Dyson_equ}
\end{equation}
where $G^{R}_0$ is the non-interacting retarded Green's function (computed using the MF solution), $G^{R}$ is the exact retarded Green's function, and $\Sigma^{R} $ is the retarded self-energy. Each of these quantities are matrix quantities in the atomically localized and spin basis and Dyson's equation has to be understood as a matrix equation. In the GW approximation, the self-energy is approximated by the (matrix) product of the Green's function and the screened potential $W$, computed within the random phase approximation (RPA), see \textit{e.g.}, Refs.~\onlinecite{joost_correlated_2019},~\onlinecite{honet_semi-empirical_2021} and~\onlinecite{honet_exact_2022} for a description of the full equations and theoretical framework. As it is common practice, we work in natural units, such that $\hbar = 1$ and $\omega$ is in energy units.

Similarly to the MF approximation, the GW approximation operates in a self-consistent manner, updating $G^{R}$ and $\Sigma^{R} $ at each step until convergence is reached for the Green's function.

\subsection{(Local) density of states, local densities and magnetic moments}

From the Green's functions, we define the spectral function: $A_{i\sigma,j\sigma'} (\omega) = -2 \Im(G^R_{i\sigma,j\sigma'}(\omega))$. The local density of states (LDOS, written $n_{i\sigma}(\omega)$) is proportional to the diagonal terms of the spectral function and the density of states (DOS, written $D(\omega)$) is the sum of all LDOS: 
\begin{equation}
n_{i\sigma}(\omega) = \frac{1}{2\pi} A_{i\sigma,i\sigma} (\omega)
\label{eq:ldos_GF}
\end{equation}
and
\begin{equation}
D(\omega) = \sum_{i\sigma} n_{i\sigma}(\omega).
\label{eq:dos_GF}
\end{equation}

The local electronic densities are found by integrating the local density of states weighted in frequency by the Fermi-Dirac statistics:
\begin{equation}
n_{i\sigma} = \int_{-\infty}^{+\infty} \dd{\omega} n_{i\sigma} (\omega) f_{FD}(\omega),
\label{eq:local_dens}
\end{equation}
where $f_{FD}(\omega)$ is the Fermi-Dirac statistics.

Finally, the local magnetic moments are defined as~\cite{joost_correlated_2019}:
\begin{equation}
\begin{split}
 \expval{ \hat{m}_i^2 } & = \expval{ (\hat{n}_{i\uparrow}-\hat{n}_{i\downarrow})^2 } \\
& =  \bigg(   n_{i\uparrow}   + n_{i\downarrow}    -  2 \hspace{0.1cm} d_i  \bigg),
\end{split} 
\label{eq:magn_mom}
\end{equation}
where $d_i = \expval{ \hat{n}_{i\uparrow} \hat{n}_{i\downarrow} } $ are the double occupancies and $\expval{ (\hat{n}_{i\sigma})^2} = \expval{ \hat{n}_{i\sigma}}  = n_{i\sigma}  $ for Fermions.

The double occupancies are found in the Green's function formalism using an adaptation of the Galitskii-Migdal formula:
\begin{equation}
    d_i = \frac{-1}{U} \sum_{k,\sigma, \bar{\sigma} } \int \frac{\dd{\omega}}{2\pi} f_{FD}(\omega - \mu) \Im \{\Sigma^{R, tot}_{i\bar{\sigma}, k\sigma} (\omega) G^R_{k\sigma, i\bar{\sigma}} (\omega) \},
\label{eq:double_occ_sig_g}
\end{equation}
where $\Sigma^{R, tot}$ is the total retarded self-energy. Since in our case the GW approximation is constructed with the MF approximation as a starting point, the retarded self-energy in eq.~(\ref{eq:Dyson_equ}) does not account for the MF self-energy, which therefore must be included in the total self-energy of eq.~(\ref{eq:double_occ_sig_g}). The total self-energy is then written 
\begin{equation}
    \Sigma^{R, tot} = \Sigma^R + \Sigma^{MF,R},
\label{eq:total_sigma}
\end{equation}
where $\Sigma^{MF,R}$ is the MF self-energy.

Splitting the double occupancies of eq.~(\ref{eq:double_occ_sig_g}) according to the total self-energy expression (eq.~(\ref{eq:total_sigma})) leads to:
\begin{equation}
    d_i = d_i^{corr} + d_i^{MF},
\label{eq:double_occ_split}
\end{equation}
where $d_i^{corr}$ (resp. $d_i^{MF}$) is the correlation (resp., MF-like) part of the double occupancies found by replacing $\Sigma^{R, tot}$ by $\Sigma^{R}$ (resp.,$\Sigma^{R, MF}$) in eq.~(\ref{eq:double_occ_sig_g}).

The MF self-energy is diagonal in spin and expressed using MF mean densities \cite{stefanucci_nonequilibrium_2013, joost_lowdins_nodate}:
\begin{equation}
\Sigma^{MF,R}_\sigma (\omega) = U \hspace{0.1 cm} \rm diag(n_{1,\bar{\sigma}}^{MF}, n_{2,\bar{\sigma}}^{MF}, \hdots, n_{N,\bar{\sigma}}^{MF} ),
\label{eq:MF_Sigma}
\end{equation}
with $\bar{\sigma}=  - \sigma$.

Using eq.~(\ref{eq:MF_Sigma}), the MF-like double occupancies of eq.~(\ref{eq:double_occ_split}) can be written as:
\begin{equation}
d_i^{MF}= \frac{1}{2} ( n^{MF}_{i,\uparrow} n_{i,\downarrow} + n^{MF}_{i,\downarrow} n_{i,\uparrow}).
\label{eq:d_iMF}
\end{equation}
    
In the MF approximation, double occupancies reduce to the MF-like ones and are given by $ d_i = d_i^{MF}=  n^{MF}_{i,\uparrow} n^{MF}_{i,\downarrow}$, leading to magnetic moments equals to:
\begin{equation}
    \expval{ (\hat{m}_i^{MF})^2 } =    n_{i\uparrow}^{MF}   + n_{i\downarrow}^{MF}    -2  n_{i\uparrow}^{MF}    n_{i\downarrow}^{MF} ,
    \label{eq:magn_mom_MF}
\end{equation}
according to eq.~(\ref{eq:magn_mom}).

\subsection{Numerical methods}

The structures were generated using the \textsc{pybinding} software~\cite{Pybinding_2020}. The numerical tools from \textsc{pybinding} package are used in Fig.~\ref{fig:1_2_EM_DOS} and in Figs. 1 and 2 of the SI. The MF and GW computations are achieved using the Hubbard$\_$GW code~\cite{honet_2023_Hubbard_GW}. A broadening parameter~\cite{honet_semi-empirical_2021} of $10^-3 \hspace{0.1cm} \frac{E_H}{t}$ (with $E_H=27.21 \hspace{0.1cm} eV$ the Hartree energy) was used for the Green's functions and the number of frequencies in the grid varied from $2^{13}$ to $2^{14}$. $2^{14}$ frequencies were needed for convergence for on-site potentials of $\epsilon =-20 \hspace{0.1cm}$ eV while $2^{13}$ was sufficient in other cases. The limits of the frequency grids were set to $\pm 16 \pi t$ except for the fully H-passivated case for which $\pm 8 \pi t$ were used.

\section{Comparison between the different methods of modeling di-hydrogenation}
\label{sec:different_modellings}

We now study the different methods (\textit{C-removing}, \textit{H-orbitals}, and \textit{C-on-sites}) presented in section~\ref{sec:methods_model_di_hydro} for modelling di-hydrogenation in the case of end-modified 7-AGNRs. As far as the \textit{H-orbitals} technique is concerned, we define effective mean density on a C atomic site hosting two H atoms as the sum of the mean densities on the C atomic site and on the additional H atom. This is further illustrated in the Supplementary Information (see Fig. 1 of the SI).

We first compare the \textit{C-on-sites} and the \textit{C-removing} methods, including TB computations similar to Ref.~\onlinecite{talirz_-surface_2017} in our comparison. In this reference, the authors showed experimentally that the BS gap is significantly reduced when the central C atoms of both ends of a 7-AGNR are di-hydrogenated. Moreover, they used TB computations and the C-removing modelling of di-hydrogenation to study the phenomenon and showed that this BS gap reduction is correctly reproduced. Table~\ref{tab:BS_gaps} reports the BS energy gaps of a finite-size 6 unit cells (UC) 7-AGNR in the TB, MF and GW approximation for the case of no end-modification as well as two end-modifications modelled using the \textit{C-removing} method and the \textit{C-on-sites} method with different on-sites values ranging from $0 \hspace{0.1 cm} eV$ to $-20 \hspace{0.1 cm} eV$. We can see in this table that the experimentally observed BS gap reduction is reproduced in all three approximations when considering the \textit{C-removing} method, reduced from $\sim 2-2.1 \hspace{0.1 cm} eV$ to $\sim 1.6-1.75 \hspace{0.1 cm} eV$. When considering the the C-on-sites method, the BS gap is also reduced and the same order of magnitude is recovered for large enough negative $\epsilon$ values.

\begin{table}
    \centering
    \begin{tabular}{|M{30mm}|M{10mm}|M{10mm}|M{10mm}|}
        \hline
    	&  \multicolumn{3}{c}{BS energy gaps [eV]} \vline \\
    	\hline
        & TB & MF & GW  \\
        \hline
        No end-modification & 1.97 & 2.12 &  2.1  \\
        C removing & 1.58 & 1.59 & 1.75  \\
        $\epsilon = 0$ eV & 1.98 & 1.99 & 2.04 \\
        $\epsilon = -4$ eV & 1.43 & 1.24 & 1.27 \\
        $\epsilon = -10$ eV & 1.56 & 1.53 & 1.69  \\
        $\epsilon = -20$ eV&  1.58 & 1.58 & 1.74  \\
        \hline
    \end{tabular}
    \caption{BS energy gaps of 6 UC 7-AGNRs in the TB, MF, and GW approximations for a non end-modified AGNR, and symmetrically modified AGNRs (both ends) using the \textit{C-removing} and \textit{C-on-sites} methods with $\epsilon$ values set to 0 eV, -4 eV, -10 eV, and -20 eV. We used $t=2.7$ eV and $U=2t$.}
    \label{tab:BS_gaps}
\end{table}

\begin{figure*}[]
\centering
    \includegraphics[width=15cm]{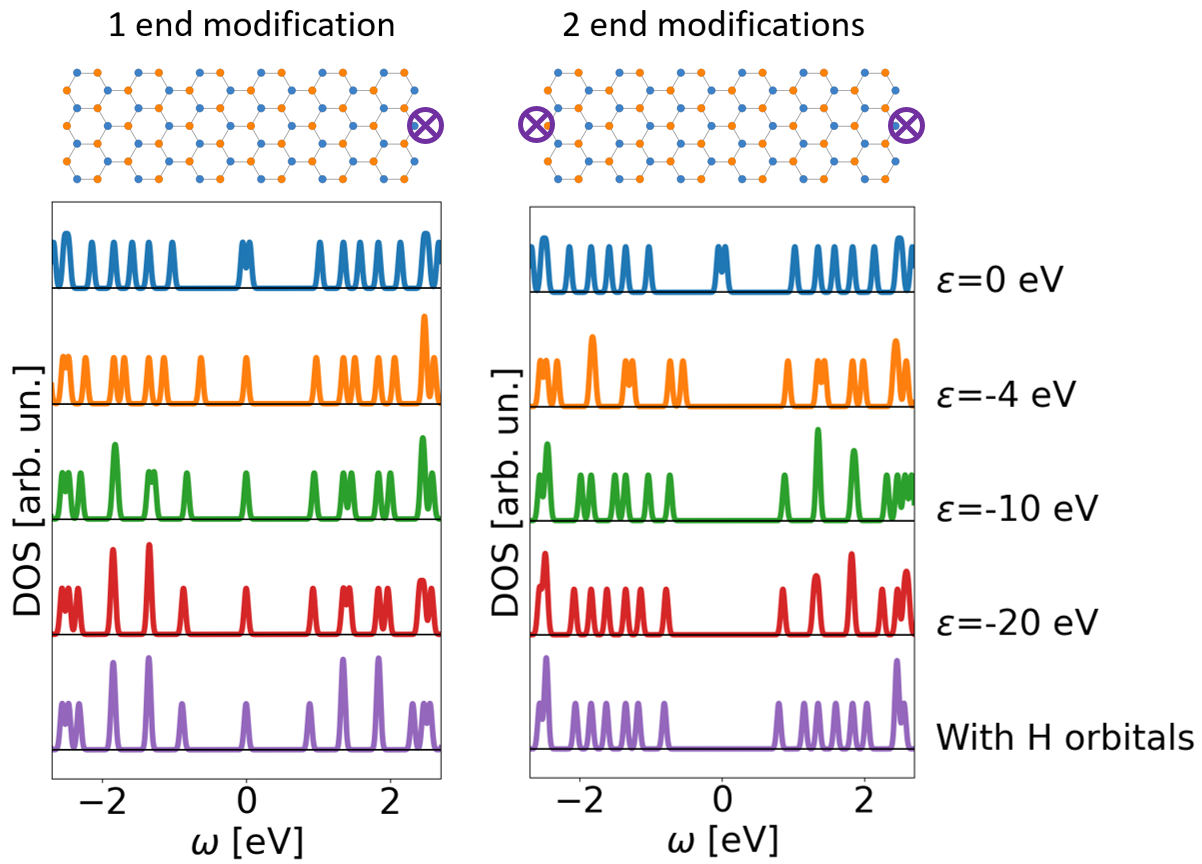}
\caption{DOS for 6 UC end-modified 7-AGNRs using the \textit{C-on-sites} method with $\epsilon$ values of 0 eV, -4 eV, -10 eV and -20 eV (top four curves) and using the \textit{H-orbitals} method (bottom curves), within a TB model (\textit{i.e.} with $U=0$) with $t=2.7$~eV. All curves are shifted artificially for better visualization and the zero DOS levels are indicated with black lines. The left (resp., right) panel shows the DOS for AGNRs with one end (resp., two ends) modified. The locations of the end modifications are indicated on the structure with purple crossed circles. The structure plots were generated using the \textsc{pybinding} software~\cite{Pybinding_2020}.  All Fermi levels are aligned to 0 eV.}
\label{fig:MF_eff_eps_DOS_wH_w2H}
\end{figure*}

Moving on to the comparison between the \textit{C-on-sites} and the \textit{H-orbitals} methods within the TB model, Fig.~\ref{fig:MF_eff_eps_DOS_wH_w2H} shows the DOS for 6 UC 7-AGNRs with one or two end(s) modified for the two methods. For the C-on-sites method, $\epsilon$ values of 0 eV, -4 eV, -10 eV and -20 eV are considered. We observe for both systems that the DOS obtained from the \textit{C-on-sites} method converges towards the \textit{H-orbitals} DOS when the magnitude of $\epsilon$ is increased. The agreement between the two methods for $\epsilon=-20$~eV in the \textit{C-on-sites} method is remarkable, especially for the unoccupied states.

\begin{figure*}
\centering
    \includegraphics[width=15cm]{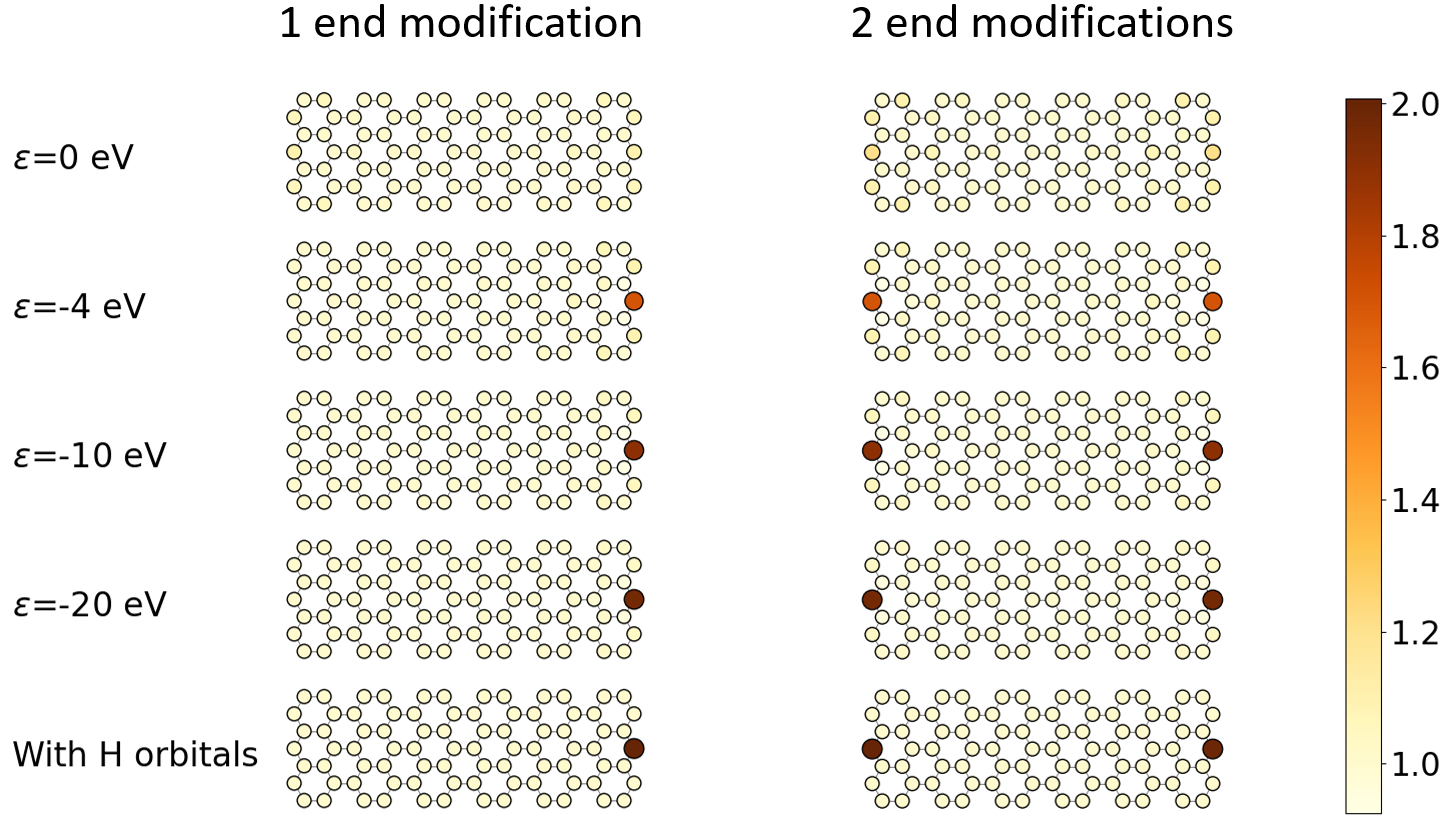}
\caption{Total electronic densities of the one-end (left) and two-end (right) modified AGNRs using the \textit{C on-site} potential and \textit{H orbitals} methods within a TB model (\textit{i.e.} with $U=0$) with $t=2.7$~eV. The \textit{C on-site} potential $\epsilon$ increases in magnitude when going downward in the panels and the last bottom panel is for the \textit{H orbital} method.}
\label{fig:dens_eff_eps_same_colorbar}
\end{figure*}

This very good agreement is confirmed by inspecting the local electronic densities of eq.~(\ref{eq:local_dens}), shown in Fig.~\ref{fig:dens_eff_eps_same_colorbar} for the same methods and parameter values. The densities are the effective densities for the \textit{H-orbitals} method as illustrated in the Supplementary Information (see Fig. 1 of the SI). As in Fig.~\ref{fig:dens_eff_eps_same_colorbar}, the scale is rather extended due to the strong localization for some models, it is instructive to also compare the electronic densities not representing the sites of strong localization for better visualization of smaller variations. This is done in the SI in Fig. 2, which allows us to also conclude that the large negative limit for $\epsilon$ reproduces the \textit{H orbitals} model well, showing a more uniform density.

In conclusion, key features of end-modified AGNRs such as the BS energy gap, the DOS, and the total electronic density can be described using the \textit{C-on-sites} method with great agreement compared to the two other modeling methods in the large enough negative value limit for the on-site potentials. Therefore, we model di-hydrogenation \textit{via} \textit{C-on-sites} method in the following of the paper, adopting a unified framework to describe di-hydrogenation and N/B substituents at the ends of AGNRs.

\section{DOS and local electronic densities}

\begin{figure}
\centering
    \includegraphics[width=8cm]{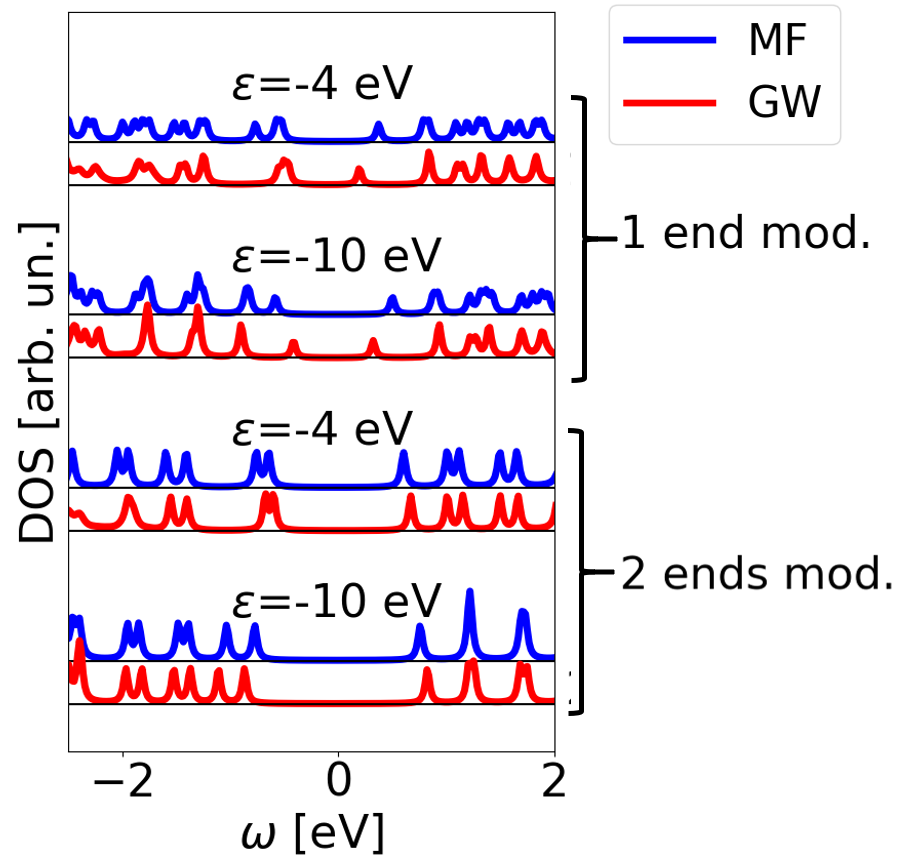}
\caption{DOS for 6 UC end-modified 7-AGNRs using \textit{C-on-sites} method with $\epsilon$ values of -4 eV and -10 eV, in the MF (blue curves) and GW (red curves) approximations. All curves are shifted artificially for better visualization and the zero DOS levels are indicated with black lines. The top (resp., bottom) four curves show DOS for AGNRs with one end (resp., two ends) modified. All Fermi levels were aligned to 0~eV. We used $t=2.7$~eV and $U=2t$.}
\label{fig:1_2_EM_DOS}
\end{figure}

For the case of finite-size 7-AGNRs with H-passivation at the edges, we showed in a previously published paper that the GW approximation introduces an energy renormalization of the topological end states and to slight changes in the total LDOS while they are more significant in the spin-polarized LDOS~\cite{honet_2023_effect}. Fig.~\ref{fig:1_2_EM_DOS} shows the DOS of end-modified 6 UC 7-AGNRs using -4 eV and -10 eV for end-modifications at one or two end(s). As for the H-passivated case, we observe little changes between the MF and GW approximations, mainly energy renormalization of near-Fermi-level states.

\begin{figure*}
\centering
    \includegraphics[width=17cm]{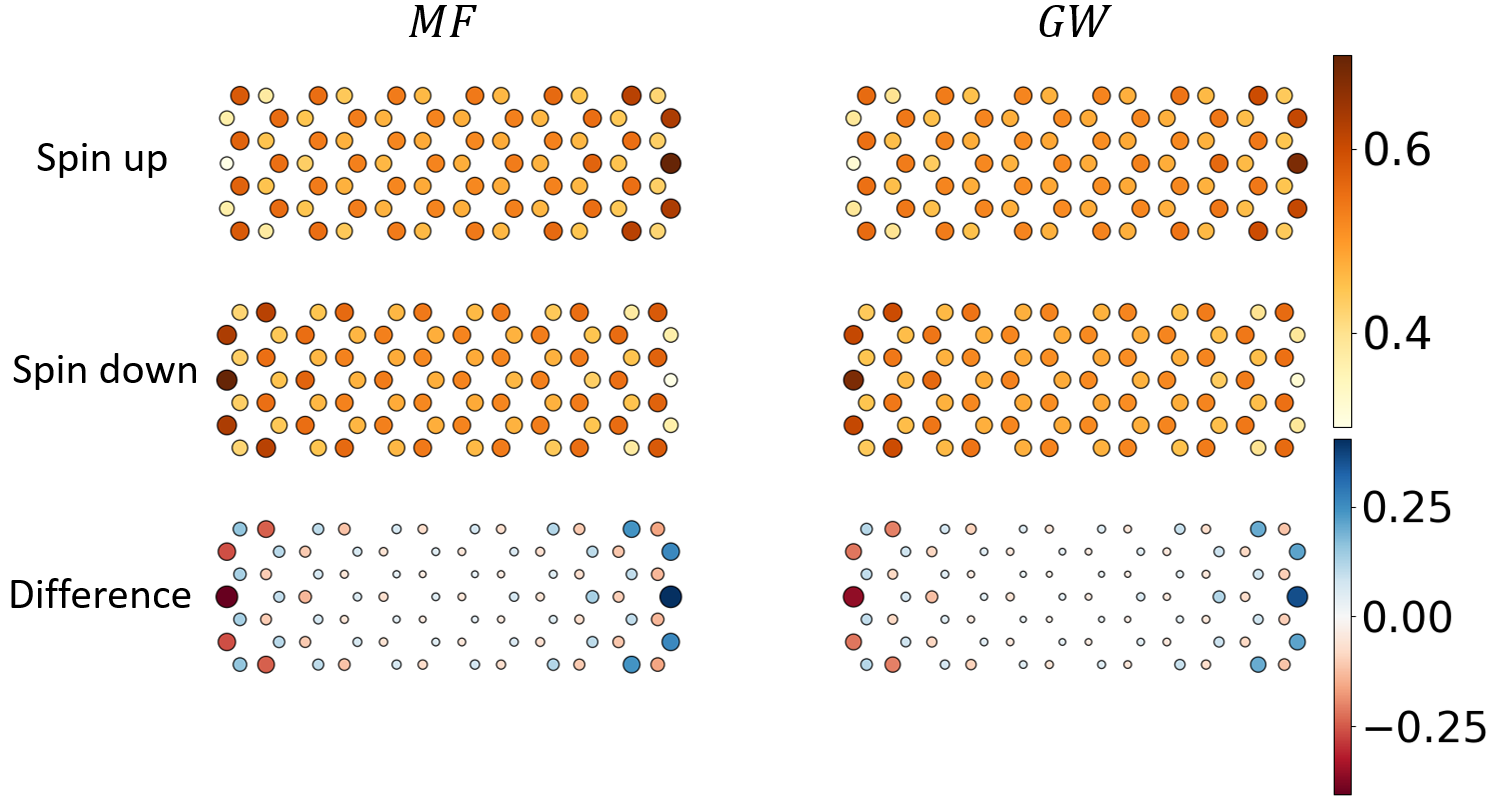}
\caption{Local electronic densities for H-passivated 6 UC 7-AGNRs. The left (resp., right) column shows the MF (resp., GW) results for spin-up and spin-down electrons (top and middle plots, respectively), as well as the difference between the two spin densities (bottom plots). We used $t=2.7$~eV and $U=2t$. }
\label{fig:total_dens_H_pass}
\end{figure*}

\begin{figure*}
\centering
    \includegraphics[width=17cm]{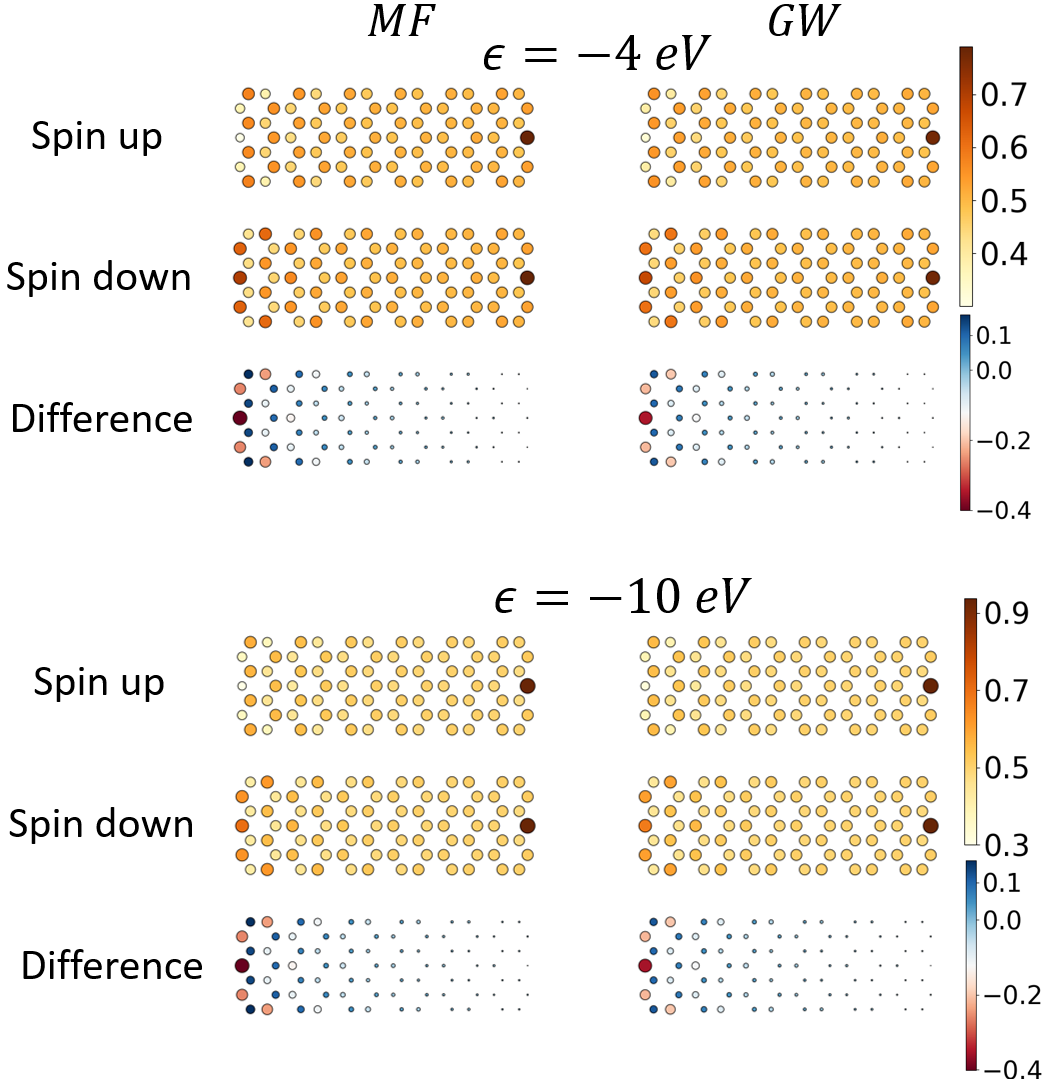}
\caption{Local electronic densities for one end-modified 6 UC 7-AGNRs. The left (resp., right) column shows the MF (resp., GW) results for spin up and spin down electrons (top and middle plots respectively) as well as the difference between the two spin densities (bottom plots). We used $t=2.7 \hspace{0.1cm}$~eV and $U=2t$}
\label{fig:total_dens_1_EM}
\end{figure*}

When incorporating an end-modification, the spin-polarization at the modified end disappears as can be seen in fig.~\ref{fig:total_dens_1_EM}. This could be understood as a consequence of the added electron occupying one more topological ES of the system. This is further illustrated in the SI in Fig. 3, where the magnetic moments of a two-electron doped system are shown, without any on-site potential. In the one-end modified case, the effect of GW approximation on the electronic density is again to reduce the spin polarization (near the unmodified end).

\begin{figure*}
\centering
    \includegraphics[width=17cm]{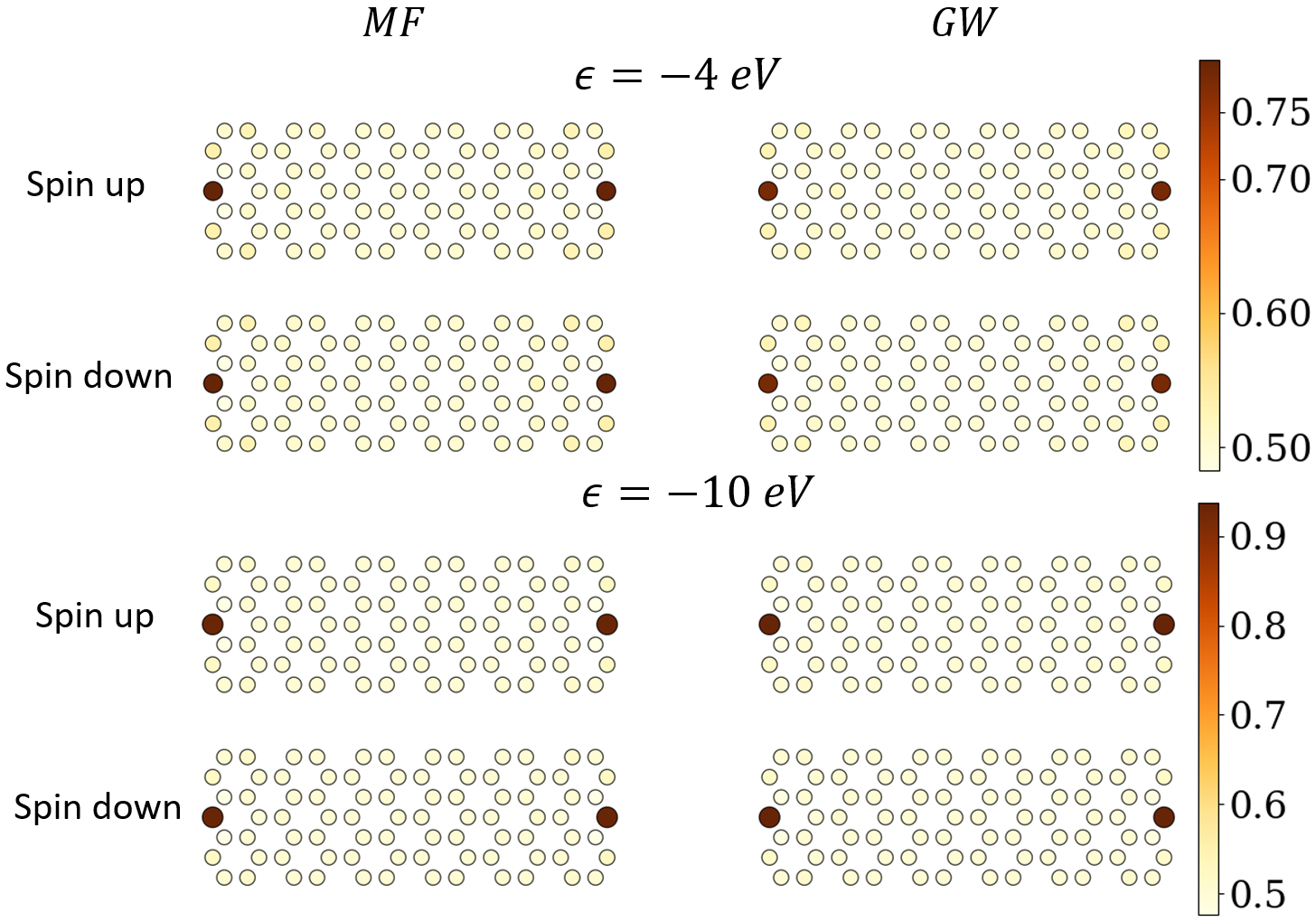}
\caption{Local electronic densities for two end-modified 6 UC 7-AGNRs. The left (resp., right) column shows MF (resp., GW) results for spin-up and spin-down electrons (top and bottom plots respectively). Compared to Fig.~\ref{fig:total_dens_1_EM}, the difference between spin-up and spin-down densities is not shown because it is zero everywhere (the densities are spin-symmetric).}
\label{fig:total_dens_2_EM}
\end{figure*}

For GNRs that are modified symmetrically at both ends, all spin polarization is removed compared to the H-passivated case, resulting in a fully spin-symmetric electron density, as can be seen in Fig.~\ref{fig:total_dens_2_EM}. As a conclusion to this section, we can state that, although there are some GW effects in the DOS and electronic densities, the MF and GW approximations lead to qualitatively similar results. GW has the effect to renormalize the energies, mostly of the topological ES and to attenuate the spin polarization of the system, but there are still opposite spin accumulations at opposite ends in the H-passivated case (see Fig.~\ref{fig:total_dens_H_pass}), a single spin accumulation at the unmodified end for the one end-modified cases (see Fig.~\ref{fig:total_dens_1_EM}) and no spin polarization for the two end-modified cases (see Fig.~\ref{fig:total_dens_2_EM}).

\section{Local magnetic moments in end-modified 7-AGNRs}
\label{sec:topol_correl_magn_mom}

Magnetic moments are quantities that are strongly affected by electronic correlation as pointed out in several studies using different methods for the inclusion of correlation~\cite{feldner_magnetism_2010, joost_correlated_2019, raczkowski_hubbard_2020}. Therefore, they are of interest for the study and quantification of correlation effects. In the MF approximation, magnetic moments can be calculated directly from mean occupations (eq.~(\ref{eq:magn_mom_MF})) while a correlated part must be included in the GW approximation.

\begin{figure*}
\centering
    \includegraphics[width=17cm]{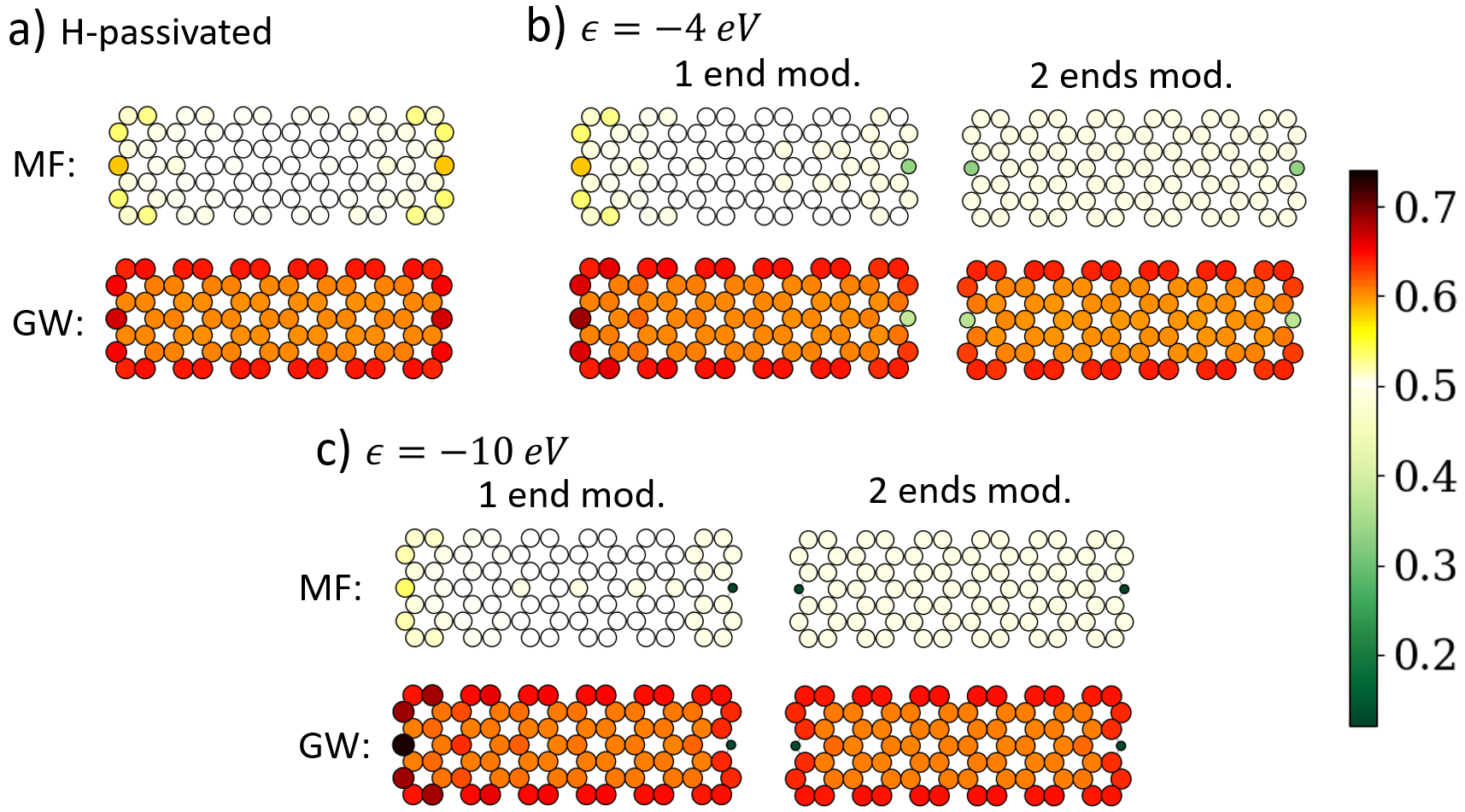}
\caption{Local magnetic moments for H-passivated (a) and for the one and two end-modified 6 UC 7-AGNRs with $\epsilon=-4 \hspace{0.1cm} eV$ (b) and $\epsilon=-10 \hspace{0.1cm} eV$ (c). In each case, the top (resp., bottom) illustrations correspond to the MF (resp., GW) results. In (b) and (c), the one (resp., two) end-modified cases are shown on the left (resp. right).}
\label{fig:magn_moments}
\end{figure*}

The magnetic moments of 6 UC 7-AGNRs computed in the MF and GW approximations are displayed in Fig.~\ref{fig:magn_moments} for the \textit{H-passivated} case in (a) and \textit{modified} cases with one and two ends using -4 eV in (b) and -10 eV in (c). For the \textit{H-passivated} case (fig.~\ref{fig:magn_moments} a)), the local magnetic moments in the MF approximation are found at the two ends where the topological ES are located. The GW approximation predicts local magnetic moments in general larger than the MF approximation. The MF magnetic moments are $\sim 0.5-0.58$ while the GW magnetic moments are $\sim 0.6-0.65$. Moreover, the GW approximation predicts the largest magnetic moments along all the edges and not only at the two zigzag ends. These observations were already made for the H-passivated case considering AGNRs heterojunctions in a recent publication using MF and GW approximations~\cite{joost_correlated_2019}. 

For the one end-modified case (see the left illustrations shown in Fig.~\ref{fig:magn_moments} b) and c)), the local magnetic moment at the site of modification decreases significantly when $\epsilon$ grows in absolute value, starting from $\sim 0.58$ (MF) and $\sim 0.67$ (GW) for the unmodified case to $\sim 0.13$ (MF) and $\sim 0.12$ (GW) when $\epsilon = -10$~eV. In the MF approximation, the magnetic moment at the opposite end (the unmodified one) decreases when $\epsilon$ increases in absolute value. Interestingly, the opposite behavior is observed in the GW approximation, resulting in a large local magnetic moment ($\sim 0.73$) at the unmodified end of the one end-modified case with $\epsilon=-10$~eV. 

For the two end-modified case (see the right illustrations in Fig.~\ref{fig:magn_moments} b) and c)), we see a decrease in the local magnetic moment at the end-modified sites, similar to the one observed in the one end-modified case. In the MF approximation, all the unmodified sites present rather uniform magnetic moments. In contrast, the GW results show stronger magnetic on the (unmodified) edges. 

Overall, while magnetic moments at the site of the modification strongly depend on the modification itself both in the MF and GW approximations, the correlated magnetic moments induced in the GW approximation (located at along all edges in the H-passivated case) appear to be robust to end modifications. They remain located at the unmodified edges and with a similar strength upon one- or two-end modifications for the different on-site potentials.

\section{Conclusion}

We conducted an investigation into the impact of end-modifications on finite-size 7-AGNRs. Our study began with a comparative analysis of various methods used in the literature to model dihydrogenation within both a tight-binding (TB) and Hubbard model frameworks. In particular, we found that adopting the \textit{C-on-site} method yielded results akin to those obtained with the \textit{C-removing} and \textit{H-orbitals} methods concerning properties such as bulk-states bandgap (BS gaps), density of states (DOS), and electronic densities.

Subsequently, with a focus on the \textit{C-on-site method}, we examined the local magnetic moments within unmodified and end-modified AGNRs. For unmodified AGNRs, our findings align with a previous study that calculated magnetic moments in GNR heterojunctions~(\cite{joost_correlated_2019}). The mean-field (MF) approximation predicted substantial magnetic moments only in regions where topological electronic structures are located, whereas the GW approximation predicted substantial magnetic moments along all edges. Additionally, we observed that edge-localized correlated magnetic moments remain robust even when end-modifications were introduced to the AGNRs, provided that the modifications were applied solely to the unaltered edges. In contrast, magnetic moments at the locations of topological electronic structures vanish when electrons were introduced into the system, leading to the occupation of previously unoccupied topological electronic structures in the case of H-passivated terminations.

These finite-size systems, synthesized experimentally, hold significant potential for future electrical and magnetic applications. However, given that they can be synthesized with various terminations, it is imperative to elucidate which properties are susceptible to termination-induced changes (\textit{e.g.}, the BS gap) and which exhibit resilience (\textit{e.g.}, local magnetic moments).


\section*{Acknowledgements}

A.H. is a Research Fellow of the Fonds de la Recherche Scientifique - FNRS. This research used resources of the "Plateforme Technologique de Calcul Intensif (PTCI)" (\url{http://www.ptci.unamur.be}) located at the University of Namur, Belgium, and of the Université catholique de Louvain (CISM/UCL) which are supported by the F.R.S.-FNRS under the convention No. 2.5020.11. The PTCI and CISM are member of the "Consortium des Équipements de Calcul Intensif (CÉCI)" (\url{http://www.ceci-hpc.be}).

\bibliographystyle{ieeetr}
\bibliography{bibliography}

\end{document}


\preprint{APS/123-QED}

\title{Robust correlated magnetic moments in end-modified graphene nanoribbons: Supporting Information}

\author{Antoine Honet}
\altaffiliation{Present address: Department of Electrical Engineering, Eindhoven University of Technology, Eindhoven 5612 AP, Netherlands}
\affiliation{%
Department of Physics and Namur Institute of Structured Materials, University of Namur, Rue de Bruxelles 51, 5000 Namur, Belgium
}%

\author{Luc Henrard}
\affiliation{%
Department of Physics and Namur Institute of Structured Materials, University of Namur, Rue de Bruxelles 51, 5000 Namur, Belgium
}%

\author{Vincent Meunier}%
\affiliation{%
Department of Engineering Science and Mechanics, The Pennsylvania State University, University Park, PA, USA
}%

\date{\today}

\maketitle


\author{Antoine Honet, Luc Henrard and Vincent Meunier}

\section{Effective representation of the local electronic densities for the H-orbitals technique}

\begin{figure*}
\centering
    \includegraphics[width=17cm]{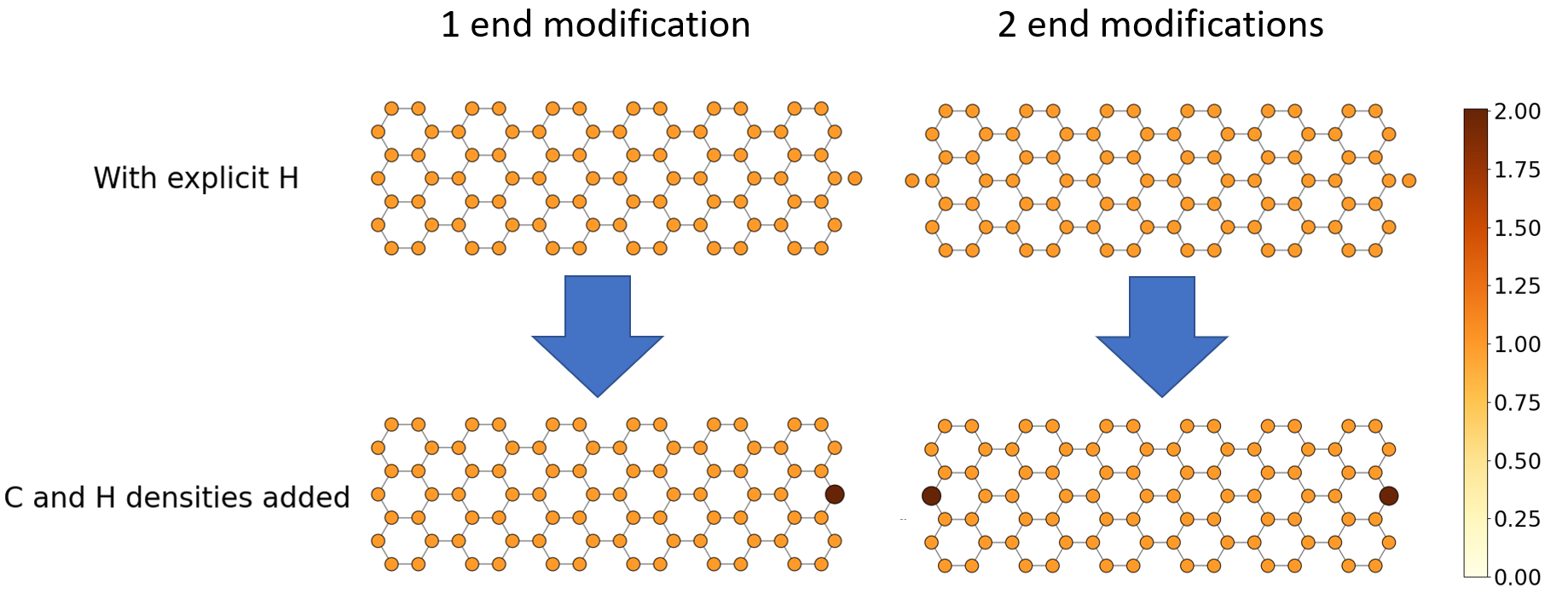}
\caption{Illustration of the choice of representation of the densities for the H-orbitals method. At the left, the AGNR is modified at one end and it is modified at the two ends for the right plots. The top panels show the densities on C atoms as well as on the added H atoms. There is one H at the right end of the left figure and there is one H atom at both ends of the right figure. The bottom panels show the chosen representation where the H densities were added to their neighbouring C atom.}
\label{fig:dens_eff_sum_arrows}
\end{figure*}

As mentioned in the main text, the H-orbitals technique introduces an H atom in addition with a pure C description of AGNRs. In order to compare with the C-on-site-potential method, we adopt a representation convention merging the densities on the added H atom and on the C atom where the H is added. This is illustrated at fig.~\ref{fig:dens_eff_sum_arrows} where we show the densities on the C atoms and the H atom separately at the top and the effective merged densities at the bottom for a one- and two-end modified 6 UC 7-AGNR.

\section{Electronic densities without the strong localised sites}

\begin{figure*}
\centering
    \includegraphics[width=17cm]{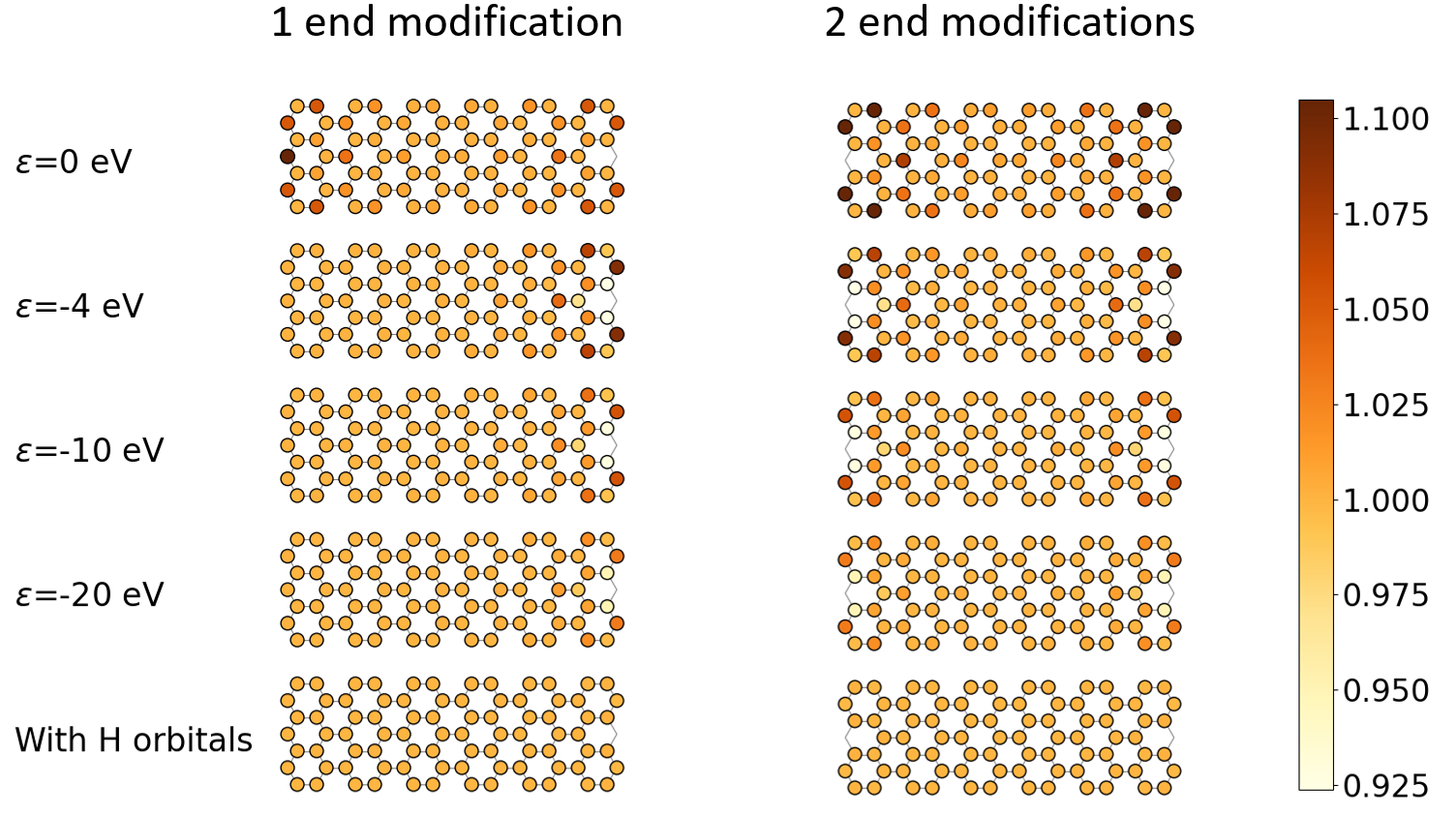}
\caption{Same as fig. 2 of the main document but not representing the sites where the strong localisation occurs and therefore adopting a different colobar scale.}
\label{fig:dens_eff_eps_delete_same_colorbar}
\end{figure*}

We provide here the local electronic densities for the comparison between C-on-site and H-orbitals methods where the end-modified sites were removed. Doing so, we adopted a different color scale to see in more details little deviations. The electronic densities are shown on fig.~\ref{fig:dens_eff_eps_delete_same_colorbar} for the one- and two-end modified cases. We see that, in addition to reproduce the strong localization on the modified sites (see main text, fig. 2), we can see that the smaller differences in the electronic densities (with an amplitude of $\sim 0.2$) are attenuated when the on-site potential increases in the C-on-site method, in agreement with the H-orbitals method showing a great uniform electronic density as far as the modified sites are not considered.

\section{Magnetic moment for the electron doping case}

\begin{figure}
\centering
    \includegraphics[width=9cm]{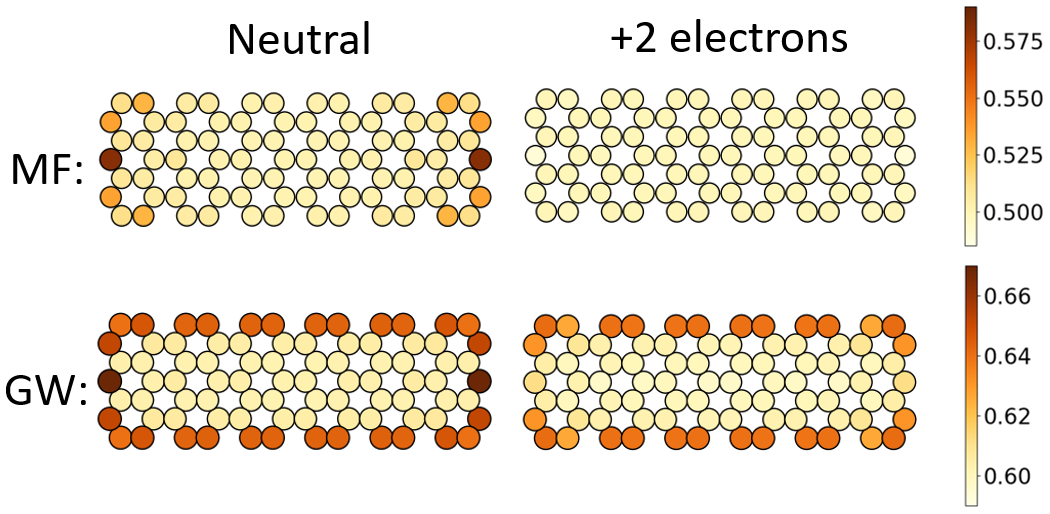}
\caption{Local magnetic moments of a 6 UC 7-AGNRs for the H-passivated case (labelled "neutral", at the left) and for the two-electron doped case (labelled +2 electrons, at the right). Top (resp. bottom) plots illustrate the MF (resp. GW) predicted magnetic moments with color scales being different. }
\label{fig:H_pass_2el_doping}
\end{figure}

We further illustrate here the effect of pure electron doping on magnetic moments in finite-size 7-AGNRs. At fig.~\ref{fig:H_pass_2el_doping}, the magnetic moments in the neutral H-passivated case and with an addition of two electrons are reported for MF and GW approximations. As stated in the main text, in the neutral case for the MF approximation the magnetic moments are located at the spatial location of the topological ES. In this case, there are two occupied and two unoccupied topological ES. Adding up two electrons to the system, we populate the four topological ES and the strong magnetic moments at the ends disappear. 

The GW approximation also predicts strong magnetic moments at the location of topological ES but additional correlated magnetic moments are predicted along all the edges. In the charged case, the magnetic moments at the ends disappear as for the MF approximation while the correlation-induced magnetic moments all the armchair edges persist. 

This case of electron doping gives a clue that the disappearance of the end localized magnetic moments is more related to the addition of electrons than to the on-site potential that is set up.

\bibliographystyle{ieeetr}
\bibliography{bibliography}